\documentclass[10pt,a4paper]{article}
\usepackage{graphicx}
\usepackage{amsmath}
\usepackage{amsthm}
\usepackage{amsfonts}
\usepackage{color}
\usepackage{amssymb}
\title{On Mean Field Limits for Dynamical Systems}

\newcommand{\A}{\mathcal{A}}
\newcommand{\B}{\mathcal{B}}
\newcommand{\C}{\mathcal{C}}
\newcommand{\D}{\mathcal{D}}
\newcommand{\M}{\mathcal{M}}
\newcommand{\F}{K}
\newcommand{\eop}{End of Proof}

\newtheorem{thm}{Theorem}[section]

\newtheorem{lem}{Lemma}[section]
\newtheorem{cor}{Corollary}[section]
\theoremstyle{definition}
\newtheorem{defn}{Definition}[section]
\theoremstyle{remark}
\newtheorem{rmk}{Remark}[section]
\newtheorem{notation}{Notation}

\author{Niklas Boers, Peter Pickl}

\begin{document}

\maketitle
\begin{abstract}
We present a purely probabilistic proof of propagation of molecular chaos for $N$-particle systems in dimension $3$ with interaction forces scaling like $1/\vert q\vert^{\lambda}$ with $\lambda<2$  and cut-off at $q = N^{-1/3}$. The proof yields a Gronwall estimate for the maximal distance between exact microscopic and approximate mean-field dynamics. This can be used to show  propagation of molecular chaos, i.e. weak convergence of the marginals to the corresponding products of solutions  of the respective mean-field equation without cut-off in a quantitative way. Our results thus lead to a derivation of the Vlasov equation from the microscopic $N$-particle dynamics with force term arbitrarily close to the physically relevant Coulomb- and gravitational forces.
\end{abstract}

\section{Introduction}

Consider a system  consisting  of $N$ interacting identical  particles  subject to Newtonian time evolution. The dynamics is given by  the respective Newtonian flow $(\Psi^N_{t,s})_{t,s\in\mathbb{R}}:\mathbb{R}^{6N}\to \mathbb{R}^{6N}$,  which is assumed to be symmetric under permutation of coordinates. Denoting the interaction force by $f:\mathbb{R}^3\to\mathbb{R}^3$ and the distribution function on phase space $\mathbb R^{6}$ with position coordinates $q$ and momentum coordinates $p$ by $k:\mathbb{R}^6\to\mathbb{R}_{0}^{+}$, the Vlasov equation is given by the non-linear PDE 
\begin{equation}
\partial_t k + \nabla_q k \cdot \dot{q} + \nabla_p k \cdot f \ast \tilde{k}_t=0\;,
\end{equation}
where $ \widetilde{k}_t(q) = \int k_{t}(q,p)d^3 p$.

The global existence and uniqueness of solutions of this equation for suitable initial conditions is well understood, even for singular interactions (see \cite{pfaffelmoser1992}, \cite{schaeffer1991} and \cite{lions1991}).

Our goal here is to derive the Vlasov equation from the microscopic Newtonian $N$-particle dynamics. For this purpose, we compare the microscopic $N$-particle time evolution with an effective one-particle description given by the Vlasov flow $(\varphi^N_{t,s})_{t,s\in\mathbb{R}}:\mathbb{R}^{6}\to \mathbb{R}^{6}$ and prove convergence of $\Psi^N_{t,s}$ to the product of $\varphi^N_{t,s}$ in the limit $N\to\infty$ in a suitable sense. From this,  weak convergence of the $s$-particle marginals of the $N$-particle system to the corresponding $s$-fold products of  solutions of the Vlasov equation follows.

This is usually referred to as \emph{propagation of molecular chaos}. Classical results of this kind are valid for Lipschitz-bounded forces \cite{Braun1977,Dob1979}. Even if formulated probabilistically, these results rely on deterministic initial conditions. Such approaches have difficulties for singular interactions in combination with clustering of particles. A very good overview is given in the book by Herbert Spohn \cite{spohn1991}.

Recently, Hauray and Jabin could include singular interaction forces scaling like $1/\vert q \vert^{\lambda}$ in three dimensions with ${\lambda} < 1$ \cite{Hauray2007} as well as the physically more interesting case with ${\lambda}$ smaller but close to $2$, and a lower bound on the cut-off at $q = N^{-1/6}$ \cite{Hauray2013}. While their convergence Theorem is deterministic as well, it is valid for quite generic initial conditions chosen according to the respective $N$-particle law. Furthermore, the latter work quantifies the rate of convergence in Wasserstein distance for large enough $N$.  

Another  deterministic   result \cite{kiessling2014}, which is valid for repulsive pair-interactions, assumes no cut-off but some additional technical condition which can be read as a  bound on the maximal forces of the microscopic system along the trajectories.  

The strategy which we shall present in the following is \emph{designed} for stochastic initial conditions. Using typicality arguments it is possible to derive the Vlasov equation where deterministic methods fail: For singular interactions there are in fact  deterministic initial conditions for which the dynamics is not  described by the Vlasov equation (see Remark \ref{cluster} below)  and in general it might be hard to rule out such initial condidtions by bounds on the energy.   However, such ``bad'' initial conditions of particles may, while not impossible, be very atypical in the sense that the respective volume in phase space is small. This offers the chance to generalize our technique and prove Vlasov-like results also for more singular potentials and/or more complicated dynamics, as for example the Vlasov-Maxwell system (for a recent result see for example \cite{Elskens2009} and \cite{Golse2012}), or for systems involving other field degrees of freedom.

In this article, the heart of the idea shall be presented for the case of forces with singularities slightly weaker than for Coulomb- or gravitational forces: $f \sim 1/\vert q \vert^{\lambda}$ with $3/2<\lambda<2$ and cut-off at $q = N^{-1/3}$. This particular cut-off width can be physically motivated by the fact that the typical inter-particle distance in position space $\mathbb{R}^3$ is given by $N^{-1/3}$. Our proof relies on a Gronwall estimate for a suitable notion of distance between exact and mean-field dynamics. We remark that the final result on the convergence of the  $s$-particle marginals to the corresponding $s$-fold products of solutions of the Vlasov equation  is quantitative in the sense that it provides the rate of convergence in $N$. 

We first state our  requirements on $\Phi^N$ and $\varphi^N$:

\begin{defn}\label{assume}
\begin{itemize}
 
\item[(a)] Let for some  $3/2<\lambda<2$ and any $N\in\mathbb{N}\cup\{\infty\}$ the interaction force $f^N:\mathbb{R}^3\to\mathbb{R}^3$ be given by 

\begin{equation}\label{eq:force}
f^N(q) = \begin{cases}a \frac{q}{|q|^{\lambda+1} } &\mbox{if } |q|\geq N^{-1/3} \\
aq N^{\frac{\lambda+1}{3}} & \mbox{else\;,} \end{cases}
\end{equation}
 where $a$ is some real number and $|\cdot |$ denotes the euclidean norm in $\mathbb R^3$ . In this sense, $f^\infty$ denotes the force without cut-off. Note that a negative $a$ corresponds to attractive forces, whereas a positive $a$ corresponds to repulsive forces.

We shall in the following use the notation $X=(Q,P)=(q_1,\ldots,q_N,p_1,\ldots,p_N)$ and $(Q)_j=q_j\in\mathbb{R}^3$. The total force of the system is given by $F:\mathbb{R}^{6N}\to\mathbb{R}^{3N}$, where $(F(X))_j:=\sum_{i\neq j} (N-1)^{-1}f^N(q_j-q_i)$ is the force exhibited on a single coordinate $j$.  Note that we omit to make the dependence of $F$ on the particle number $N$ explicit. 

\item[(b)]  Let $\Psi^N_{t,s}$ be the Newtonian flow on $\mathbb{R}^{6N}$, defined by
\begin{equation}\label{diff}\frac{d}{dt}\Psi^N_{t,s}(X)=V(\Psi^N_{t,s}(X))
\end{equation}
where $V$ is given by $V(X)=(P,F(X))$. 

\item[(c)]
We introduce now for any probability density $k_0:\mathbb{R}^{6}\to \mathbb{R}_{0}^{+}$ the effective one particle flow $(\varphi^N_{t, s})_{t\geq s}$ given by the following coupled equations: First, we define $k:\mathbb{R}\times\mathbb{R}^{6}\to \mathbb{R}^{+}_0$, which gives for each time $t$ the effective distribution function time-evolved with respect to $\varphi^N_{t, s}(x)$ : $k(0,\cdot)=k_0$ and 
\begin{align}
\label{notf}
k^N_t(x):=k^N(t,x)=k_0(\varphi^N_{0,t}(x))\;.
\end{align}
Second, for $x = (q, p)$, the effective flow $\varphi^N_{t, s}$ itself is defined by
\begin{align}\label{vlasov}
\frac{d}{dt} \varphi^N_{t, s}(x)=&v^t(\varphi^N_{t, s}(x))\;,
\end{align}
where $v^t$ is given by $v ^t(x) = (p, \overline{f}^N_t(q))$. Here, the \emph{mean-field force} $\overline{f}^N_t$ is defined as $\overline{f}^N_t=f^N \ast \widetilde k^N_t$ and $\widetilde{k}^N:\space\mathbb{R}\times\mathbb{R}^3\to \mathbb{R}^+_0$ is given by
$$\widetilde{k}^N_t(q):=\int k^N_t(q,p)d^3p\;.$$ 
 
\item[(d)]

We shall  lift this flow to the $N$-particle  phase-space. The respective $\Phi^N_{t, s}  = \left(\varphi_{t,s}^N\right)^{\otimes N}$ satisfies 
\begin{align*}
\frac{d}{dt}\Phi^N_{t, s}(X)=\overline V_t(\Phi^N_{t, s}(X))\;,
\end{align*}
with $\overline V_t(X) = (P,\overline F_t(Q))$ and $\overline F_t$ given by $\left(\overline F_t(Q)\right)_j:=\overline{f}^N_t(q_j)\;.$

\end{itemize}

\end{defn}

\begin{rmk}
It is possible to generalize to $0\leq\lambda\leq3/2$  and to dimension $d\neq3$. Depending on $\lambda$  and $d$, it is furthermore possible to choose a narrower cut-off.  To keep the notation as simple as possible, we restrict ourselves to the situation above.
\end{rmk}

\begin{rmk}\label{cluster}
In the model we are considering there are in fact configurations $X$ for which the Wasserstein distance between the empirical densities and the effective distribution function will, despite being initially small, become large during time evolution. Consider a configuration $X$ for which groups of $N^ {3/4}$ particles cluster in the sense that they are all located at the same coordinate. There are $N^ {1/4}$ such clusters in total, and we assume that each of them is distributed independently according to a probability density $k$. Choose a typical distribution $\delta$ according to the law $\prod^{N^{1/4}}k$. Then the initial Wasserstein distance between the empirical distribution $\delta$ and the effective distribution $k$ is small. However, the potential energy of each particle-pair in a given cluster scales as $N^{-2/3}$, and thus the potential energy of each of the $N$ particles diverges as $N\to\infty$. Eventually, the clusters will break up, resulting in large deviations
  of the kinetic energies and consequently the momenta from the mean-field case.
\end{rmk}

\begin{notation}

$k^N_t:\mathbb{R}^6\to\mathbb{R}^+_0$ can be understood as a one particle probability density. All probabilities and expectation values are meant with respect to the product measure given at a certain time, i.e. for any random variable $H:\mathbb{R}^{6N}\to\mathbb{R}$ and any  element $A$ of the  Borel $\sigma$-algebra  
\begin{align}
\mathbb{P}_t(H\in A)=& \int_{H^{-1}(A)} \prod_{j=1}^N k^N_t(x_j)dX\\
\mathbb{E}_t(H)=& \int_{\mathbb{R}^{6N}} H(X)\prod_{j=1}^N k^N_t(x_j)dX\;.
\end{align}

\end{notation}
Since the  measure is invariant under $\Phi^N_{t, s}$,  it follows that
\begin{align*}
\mathbb{E}_s(H\circ\Phi^N_{t, s})
=&
\int_{\mathbb{R}^{6N}} H(\Phi^N_{t, s}(X))\prod_{j=1}^N k^N_s(x_j)dX
\\
=&\int_{\mathbb{R}^{6N}} H(X)\prod_{j=1}^N k^N_s(\varphi^N_{s,t}(x_j))dX\;.
\end{align*}
Since $k^N_s(\varphi^N_{s,t}(x_j))=k^N_t(x_j)$ it follows that
\begin{align}
\mathbb{E}_s(H\circ\Phi^N_{t, s})=\mathbb{E}_t(H)\label{zeitumw}\;.
\end{align}

 We first state our result for interaction forces with cut-off:
 
\begin{thm}\label{theo1}
Let $t>0$  be such that there exists a  $C_0<\infty$  with
\begin{equation}\label{cond}
\sup_{0\leq s\leq t}\|\widetilde{k}^N_s\|_\infty<C_0\;.
\end{equation} 
Then, under the assumptions given in Definition \ref{assume}, there exists a constant $C_1<\infty$  such that 
\begin{align}\label{maineq}
\mathbb{P}_0(\sup_{0\leq s\leq t}\left|\Psi^N_{s,0}(X)-\Phi^N_{s,0}(X)\right|_\infty>N^{-1/3})\leq C_1 N^{\frac{2\lambda-4}{9}}\;.
\end{align}

\end{thm}

Under further assumptions on the initial conditions $k_0$, as well as on the solution of the Vlasov equation \emph{without} cut-off (denoted by $k_t^\infty$), we can show that solutions for forces with cut-off approximate solutions for forces without cut-off. Consequently, under these additional assumptions, Theorem \ref{theo1} can be generalized to the case of interaction forces without cut-off:

 \begin{thm}\label{theo2}
Let  $\nabla k_0(x)\leq C_0(1+|x|)^{-7}$  for some $C_0<\infty$.
 Let $t>0$  be such that  there exists a  $C_1<\infty$   with
\begin{equation}\label{cond2}
 \sup_{0\leq s\leq t}\|\widetilde{k}^\infty_s\|_\infty<C_1\;.
 \end{equation}  
 Then, under the assumptions given in Definition \ref{assume}, there exist constants $C_2,C_3<\infty$ such that 
\begin{align}\label{maineq2}
\mathbb{P}_0(\sup_{0\leq s\leq t}\left|\Psi^N_{s,0}(X)-\Phi^\infty_{s,0}(X)\right|_\infty>C_2 N^{-1/3})\leq C_3 N^{\frac{2 \lambda-4}{9}}\;.
\end{align}
\end{thm}

We can further show that Theorem \ref{theo1} (respectively Theorem \ref{theo2}) implies weak convergence of the  $s$-particle marginals of the microscopic $N$-particle system to the corresponding products of solutions  of the Vlasov equation with cut-off, i.e. $k^N_t$ (respectively without cut-off, i.e. $k^\infty_t$):

\begin{defn}
Let $\mathcal{L}$ be the space of functions $f: \mathbb{R}^6\to\mathbb{R}$ given by
$$f\in\mathcal{L}\Leftrightarrow \|f\|_\infty=\|f\|_L=1\;,$$
where $\|f\|_L$ denotes the global Lipschitz constant of $f$.

 For two probability densities $k,l:\mathbb{R}^6\to\mathbb{R}^+_0$ the \emph{bounded Lipschitz distance} is defined by
$$d_L(k,l):=\sup_{f\in \mathcal{L}}\left|\int (k(x)-l(x))f(x)d^6x\right|\;.$$ 

\end{defn}

\begin{cor} \label{cor1}

Let $t>0$, and let for probability densities $k_0: \mathbb{R}^6\to\mathbb{R}^+_0$ the $N$-particle densities $\F:\mathbb{R}\times \mathbb{R}^{6N}\to\mathbb{R}^+$ be given by $\F_0(X)=\prod_{j=1}^N k_0(x_j)$ and $\F_t(X):=\F_0(\Psi^N_{0,t}(X))$.

Then, under the conditions of Theorem \ref{theo1} (respectively Theorem \ref{theo2}), the reduced  $s$-particle marginal given by 
$$\F^{(s)}_t(x_1, \dots, x_s):=\int \F_t(X) d^6x_{s+1}d^6x_{s+2}\ldots d^6x_N$$ 
converges weakly to the $s$-fold product of $k^N_t$ (respectively $k^\infty_t$) in the sense that 

\begin{align*}
d_L(\F^{(s)}_t,\left(k^N_t\right)^{\otimes s})\leq   CN^{\frac{2 \lambda-4}{9}} 
\;\;\;\;\;\;\;\;
\left(\text{respectively }\;\;d_L(\F^{(s)}_t,\left(k^\infty_t\right)^{\otimes s})\leq   CN^{\frac{2 \lambda-4}{9}}  \right)
\end{align*}

for some $C<\infty$.

\end{cor}

\section{Proof of Theorem \ref{theo1}}\label{s2}

\begin{notation}
Constants appearing in estimates will generically be denoted by $C$. We shall not distinguish constants
 appearing in a sequence of estimates, i.e. in $X\leq  CY\leq  CZ$,  the constants $C$ may differ. 
\end{notation}

We first introduce a suitable notion of distance on $\mathbb{R}^{3N}$ which enables us to prove that for finite time  $\Psi^N_{t,0}$ and $\Phi^N_{t, 0}$ will typically be close with respect to that notion of distance. Since we are dealing with probabilistic initial conditions, we introduce a stochastic process $J_t$ which is such that 
\begin{itemize}
\item[a)] we can show that the expectation value of $J_t$ is small and
\item[b)] a small expectation value of $J_t$ implies that - typically - $\Psi^N_{t,0}(X)$ and $\Phi^N_{t, 0}(X)$ are close, i.e. Theorem \ref{theo1}.  
\end{itemize}

\begin{defn}
Let  $J_t:\mathbb{R}^{6N}\times\mathbb{R}\to\mathbb{R}$   be the stochastic process given by 
$$J_t(X):=\min\left\{1,N^{1/3}\sup_{0\leq s\leq t}\left|\Psi^N_{s,0}(X)-\Phi^N_{s,0}(X)\right|_\infty\right\}\;.$$

Here $|\cdot|_\infty$ denotes the supremum norm on $\mathbb{R}^{6N}$.
\end{defn}

We shall prove the following

\begin{lem}\label{mainlem}
Let $t>0$. Then, under the assumptions of Theorem \ref{theo1}, there exists a constant $C<\infty$ such that
\begin{align}
\mathbb{E}_0(J_t)\leq C N^{\frac{\lambda-4}{9}} \;.
\end{align}

\end{lem}

Since the probability $$\mathbb{P}_0\left(\sup_{0\leq s\leq t}\left|\Psi^N_{s,0}(X)-\Phi^N_{s,0}(X)\right|_\infty\geq N^{-1/3}\right)=\mathbb{P}_0(J_t=1)\leq \mathbb{E}_0(J_t)\;, $$ Theorem \ref{theo1} is a direct consequence of the Lemma.  The proof of Lemma \ref{mainlem} is given at the very end of section \ref{s2}.

The proof of the Lemma need some further preparation. It is based on a Gronwall argument, i.e. we shall give an upper bound on the difference 
\begin{equation}
\label{groe}
\mathbb{E}_0(J_{t+dt}-J_{t})=\mathbb{E}_0(J_{t+dt})- \mathbb{E}_0(J_{t})\;.\end{equation}

We will do so by a suitable partition of the phase space $\mathbb{R}^{6N}$.
\begin{defn}
For any subset $\A\subset\mathbb{R}^{6N}$, any random variable $J$ and any $s$ we define the restricted expectation value of $J$ with respect to $k^N_s$ by
$$\mathbb{E}_s(J\mid \A):=\mathbb{E}_s(J^\A)$$
where $J^\A$ is the random variable given by
$$J^\A(X)=
\begin{cases}
J(X) & \text{ if }X\in \A \\
0 & \text{ else.}
\end{cases}$$

\end{defn}

From this definition it follows directly that
\begin{align}
 \mathbb{E}_0(J)=\sum_{j\in I}  \mathbb{E}_0(J\mid \A_j)
\end{align}
for any partition $\dot{\bigcup}_{j\in I}A_j=\mathbb{R}^{6N}$.

Our strategy can now be summarised as follows: First note that those configurations where $J_t$ is maximal, i.e. $\left|\Psi^N_{t,0}(X)-\Phi^N_{t,0}(X)\right|_\infty\geq N^{-1/3}$, are irrelevant for finding an upper bound of $\mathbb{E}_0(J_{t+dt})-\mathbb{E}_0(J_t)$. Below we shall call the set of such configurations $\A_t$ and show that $\mathbb{E}_0(J_{t+dt}-J_{t}\mid \A_t)\leq 0$. 

We are thus left with configurations for which $\left|\Psi^N_{s,0}(X)-\Phi^N_{s,0}(X)\right|_\infty< N^{-1/3}$.  
The growth of $\mathbb{E}_0(J_{t+dt})-\mathbb{E}_0(J_t)$ stems from fluctuations in the forces: 
\begin{align*}
\left|F(\Psi^N_{t,0}(X))-\overline F(\Phi^N_{t,0}(X))\right|_{\infty}&\leq \left|F(\Psi^N_{t,0}(X))- F(\Phi^N_{t,0}(X))\right|_{\infty}\\&+\left|F(\Phi^N_{t,0}(X))-\overline F(\Phi^N_{t,0}(X))\right|_{\infty}\;.
\end{align*}
Note that $\Phi^N_{t,0}(X)$ is product distributed.
The set of configurations $X$  for which the first term $\left|F(\Phi^N_{t,0}(X))-\overline F(\Phi^N_{t,0}(X))\right|_{\infty}$ is large will be denoted by $\B_t$. Large means in our case ``larger than $N^{-1/3}$'', since any difference in the force is directly translated into a growth in the difference $\left|\Psi^N_{t,0}(X)-\Phi^N_{t,0}(X)\right|_\infty$, which is multiplied by $N^{1/3}$ in the definition of $J_t$. We shall show below that the probability to be in $\B_t$ is small.

If $F$ was globally Lipschitz continuous, the  second term  $\left|F(\Phi^N_{s,0}(X))-F(\Psi^N_{s,0}(X))\right|_\infty$ would directly translate into the difference $\left|\Psi^N_{s,0}(X)-\Phi^N_{s,0}(X)\right|_\infty$ and the result would be proven. However, the forces we consider are mildly singular, and hence there is no uniform Lipschitz constant (in particular not in $N$).  There exist in fact configurations $X$ - for example when all particles have the same position - for which this force becomes singular in the limit $N\to\infty$. However, we shall show that, for {\it typical} distributions, the force does indeed satisfy a Lipschitz condition. 

To implement this argument, we will introduce a function $g$ which can be used to control the difference $|f^N(x)-f^N(x+\delta)| _\infty$. Here, $\delta$ is smaller than $2 N^{-1/3}$, since in the coordinates of any two interacting particles we only need to take into account fluctuations smaller than $N^{-1/3}$. We will then control $G=(N-1)^{-1}\sum_{j\neq k}g(q_j-q_k)$ for typical configurations. Below we will denote the set of configurations where $G$ is large by $\C_t$ and show that it is very improbable to be in $\C_t$. For the configurations which are left we have a Lipschitz condition on $F$ and will get a good estimate on $\left|F(\Phi^N_{t,0}(X))-F(\Psi^N_{t,0}(X))\right|_\infty$.

\begin{defn}
Let $$g(q):= \begin{cases} \frac{94}{|q|^{\lambda+1}} &\mbox{if } |q|\geq 3\sqrt{3}N^{-1/3}  \\
 \sqrt{3}N^{\frac{\lambda+1}{3}} & \mbox{else. } \end{cases}$$
and 
 $G$ be defined by $(G(X))_j:=\sum_{j\neq i}(N-1)^{-1} g(q_j-q_i)$. Furthermore let $\overline{G}_t$ be  given by $\left(\overline{G}_t(X)\right)_j:= \overline{g}_t(q_j)$ with $\overline{g}_t(q)=  g\ast \widetilde k^N_t(q)$.

\end{defn}

\begin{lem}\label{gkraft}
For any $\delta\in\mathbb{R}^3$ with $|\delta|_\infty< 2N^{-\frac{1}{3}}$ it follows that
$$|f^{N}(q)-f^{N}(q+\delta)|_\infty\leq g(q) |\delta|_\infty \;.$$
\end{lem}

\begin{proof}
First note that the derivative of $f^{N}$ is bounded by $N^{\frac{\lambda+1}{3}}$. Hence for $|q| < 3\sqrt{3}N^{-1/3}$ using that $|\cdot|\leq\sqrt{3}|\cdot|_\infty$
\begin{align*}|f^{N}(q)-f^{N}(q+\delta)|_\infty&\leq |f^{N}(q)-f^{N}(q+\delta)|\leq N^{\frac{\lambda+1}{3}}|\delta| \leq \sqrt{3}N^{\frac{\lambda+1}{3}}|\delta|_\infty 
\\&=g(q) |\delta|_\infty
\end{align*}

For $|q|\geq 3\sqrt{3}N^{-1/3}$ we have the largest difference if $\delta$ points in the opposite direction of $q$. The largest derivative between $q$ and $q+\delta$ is then at the point closest to the center. It follows that
\begin{align}
|f^{N}(q)-f^{N}(q+\delta)|_\infty\leq& 
\left|\frac{d}{dr}(r^{-\lambda})\big|_{r=|q|-|\delta|}\right| \left|\delta\right|
\\=& \lambda(|q|-|\delta|)^{-\lambda-1}\left|\delta\right|\;.
\end{align} 
Since  $\frac{2}{3}|q|\geq2\sqrt{3}N^{-1/3}>\sqrt{3} |\delta|_\infty\geq |\delta|$ it follows that $|q|-|\delta|=\frac{|q|}{3}+\frac{2}{3}|q|-|\delta|>\frac{|q|}{3}$

Using this, $\lambda<2$ and $|\cdot|\leq\sqrt{3}|\cdot|_\infty$, we get
\begin{align}
|f^{N}(q)-f^{N}(q+\delta)|_\infty<& \lambda 3^{\lambda+1}|q|^{-\lambda-1}\sqrt{3}|\delta|_\infty< 94 |q|^{-\lambda-1}|\delta|_\infty\;.
\end{align}

\end{proof}

\begin{defn}\label{defsets} Let for any $t\in\mathbb{R}$ the sets $\A_t,\B_t,\C_t\subset\mathbb{R}^{6N}$ be given by 
\begin{align}
X\in \A_t&\Leftrightarrow |J_t(X)|=1\\
X\in \B_t&\Leftrightarrow |F(\Phi^N_{t,0}(X))-\overline F_t(\Phi^N_{t,0}(X))|_\infty>N^{\frac{2\lambda-7}{9}}\\
X\in \C_t&\Leftrightarrow |G(\Phi^N_{t,0}(X))-\overline G_t(\Phi^N_{t,0}(X))|_\infty>1\;.
\end{align}
\end{defn}

Since $J_{t}$ is bounded by one and $J_{t}(X)=1$ for any $X\in \A_t$ it follows that 
$J_{t+dt}(X)-J_{t}(X)\leq 0$ for any $X\in \A_t$. Therefore
\begin{equation}
\label{est1}
\mathbb{E}_0(J_{t+dt}-J_{t}\mid \A_t)\leq 0\;.
\end{equation}

By definition \ref{assume} it follows that 
\begin{align}
\Psi^N_{t+dt,0}(X)=\Psi^N_{t,0}(X)+V(\Psi^N_{t,0}(X))dt +o_N(dt)\\
\Phi^N_{t+dt,0}(X)=\Phi^N_{t,0}(X)+\overline V_t(\Phi^N_{t,0}(X))dt +o_N(dt)
\end{align}
where the index $N$ on $o_N$ appears in order to remind the reader that the limit is {\it not} uniform in $N$.

By triangle inequality we get that
\begin{align*}\left|\Psi^N_{t+dt,0}(X)-\Phi^N_{t+dt,0}(X)\right|_\infty\leq&\left|\Psi^N_{t,0}(X)-\Phi^N_{t,0}(X)\right|_\infty\\&+\left|V(\Psi^N_{t,0}(X))-\overline V_t(\Phi^N_{t,0}(X))\right|_\infty dt+o_N(dt)\;,
\end{align*}
therefore
$$J_{t+dt}(X)-J_{t}(X)\leq 
 \left|V(\Psi^N_{t,0}(X))dt-\overline V_t(\Phi^N_{t,0}(X))\right|_\infty N^{1/3}dt +o_N(dt)
 $$
and thus

\begin{equation}
\label{est2}
\mathbb{E}_0(J_{t+dt}-J_{t}\mid \A_t^c)-\mathbb{E}_0(\left|V\circ\Psi^N_{t,0}-\overline V_t\circ\Phi^N_{t,0}\right|_\infty\mid \A_t^c)N^{1/3}dt=o_N(dt)\;,
\end{equation}
where $\A_t^c$ denotes the complement of the set $\A_t$.
 
We are left with estimating $\mathbb{E}_0(V\circ\Psi^N_{t,0}-\overline{V}_t\circ\Phi^N_{t,0}\mid \A_t^c)$ and will now further partition the set $\A_t^c$ using definition \ref{defsets}:

\begin{align}
\mathbb{E}_0&\left(J_{t+dt}-J_{t}) =\mathbb{E}_0(J_{t+dt}-J_{t}\mid \A_t\right) \tag{$\leq 0$}\nonumber
\\&+\mathbb{E}_0\left(J_{t+dt}-J_{t}\mid \A_t^c\right)-\mathbb{E}_0\left(\left|V(\Psi^N_{t,0}(X))-\overline V_t(\Phi^N_{t,0}(X))\right|_\infty \mid \A_t^c\right)N^{1/3}dt \tag{$=o_N(dt)$}\nonumber
\\&+\mathbb{E}_0\left(\left|V(\Psi^N_{t,0}(X))-\overline V_t(\Phi^N_{t,0}(X))\right|_\infty \mid (\B_t\cup \C_t)\backslash \A_t\right)N^{1/3}dt\nonumber
\\&+\mathbb{E}_0\left(\left|V(\Psi^N_{t,0}(X))-\overline V_t(\Phi^N_{t,0}(X))\right|_\infty  \mid (\B_t\cup \C_t\cup \A_t)^c\right) N^{1/3}dt\nonumber\;.
\\\leq&\left(\sup_{X\in\mathbb{R}^{6N}}\left\{|F(X)|_\infty\right\}+\sup_{X\in\mathbb{R}^{6N}}\left\{ |\overline F(X)|_\infty\right\} + N^{-1/3}\right)\left(\mathbb{P}_0(\B_t)+\mathbb{P}_0(\C_t)\right)N^{1/3}dt\nonumber 
\\+&\mathbb{E}_0\left(\left|V(\Psi^N_{t,0}(X))-\overline {V}_t(\Phi^N_{t,0}(X))\right|_\infty  \mid (\B_t\cup \C_t\cup \A_t)^c\right) N^{1/3}dt\nonumber
+o_N(dt)\;,
\end{align}

where we used that on the set $(\B_t \cup \C_t)\backslash \A_t$, the momentum part of $\left|V(\Psi^N_{t,0}(X))-\overline V_t(\Phi^N_{t,0}(X))\right|$ is bounded by $N^{-1/3}$.

Since the two-particle force is bounded by $N^{2/3}$ it follows that the total force acting on each particle is also bounded by $N^{2/3}$.  Furthermore, the mean-field force is bounded, thus 
\begin{align}
\mathbb{E}_0(J_{t+dt}-J_{t}) \leq& (2N + 1) \left(\mathbb{P}_0(\B_t)+\mathbb{P}_0(\C_t)\right) dt\label{line1}
\\&+\mathbb{E}_0\left(\left|V(\Psi^N_t(X))-\overline V_t(\Phi^N_t(X))\right|_\infty  \mid (\B_t\cup \C_t\cup \A_t)^c \right) N^{1/3}dt\label{line2}
\\&+o_N(dt)\;.\nonumber
\end{align}

At this point we have arrived at the crucial estimates of our proof. Based on the law of large numbers, we shall now show that the probability to be in the set $\mathcal B$ or $\mathcal C$ is smaller than $C_\gamma N^{-\gamma}$ for any $\gamma>0$ and some $C_\gamma$.  This yields that (\ref{line1}) is small (see Corollary \ref{largenumbers}). The control of (\ref{line2}) is then provided in Lemma \ref{lipschitz}.

Since $f$ and $g$ do have some similarities, we shall give the law of large numbers argument for an appropriate general function $h$ and use this general estimate thereafter to control $\mathbb{P}_0(\B_t)$ and $\mathbb{P}_0(\C_t)$ in Corollary \ref{largenumbers}.

\begin{lem}\label{largenumberslem}
Let  $a\in\mathbb{N}$ and $h:\mathbb{R}^3\to\mathbb{R}^a$  be a function with $$|h(q)|\leq\begin{cases} C N^{-\frac{2\lambda+2}{9}}|q|^{-\lambda}&\mbox{if } |q|\geq N^{-1/3} \\CN^{\frac{\lambda-2}{9}}& \mbox{else\;,} \end{cases}\;,$$

for some $C\in\mathbb{R}^+$.
Let $H_j(X):=\sum_{i\neq j} h(q_j-q_i)$ and $\D_j\subset\mathbb{R}^{6N}$ be given by
$$X\in \D_j\Leftrightarrow \left|H_j(X)- (N-1)h\ast \widetilde k^N_t(q_j)\right|>1$$ with $\widetilde k^N_t$ as in Theorem \ref{theo1}
and
$$\D=\bigcup_{j=1}^N \D_j\;.$$

Then  there exists a $C_\gamma<\infty$ for any $\gamma>0$ such that
$\mathbb{P}_t(\D)\leq C_\gamma N^{-\gamma}$.
\end{lem}

\begin{proof}
Due to symmetry in exchanging any two coordinates $$\mathbb{P}_t(\D)\leq \sum_{j=1}^N\mathbb{P}_t(\D_j)=N\mathbb{P}_t(\D_1)\;.$$
So it is sufficient to show that for any $\gamma>0$ there exists a $C_\gamma<\infty$ such that
\begin{equation}
\label{daszeigen} \mathbb{P}_t(\D_1)\leq  C_\gamma N^{-\gamma}\;.
\end{equation}

The proof of (\ref{daszeigen}) is based on a law of large numbers argument. Using Markov we get that for any even natural number $M$ 
$$\mathbb{P}_t(\D_1)\leq \mathbb{E}_t\left(\left(H_1(X)-(N-1)h\ast \widetilde{k}^N_t(q_1)\right)^M\right)\;.$$

So let  $M\in2\mathbb{N}$ be some even natural number. 

Let $\mathcal{M}$  be a set of multi-indices, more precisely   the set of all maps $\alpha:\{1,2,\ldots,M\}\to\{2,\ldots,N\}$. Define $|\cdot|:\mathcal{M}\to\mathbb{N}$ as the number of elements in the image of $\alpha$
$$|\alpha|=|\alpha(\{1,\ldots,M\})|\;.$$
For any $1\leq j\leq N$ let $\alpha_j:=\sum_{i=1}^M \delta(\alpha(i),j)$ and
\begin{equation}\label{defG}
G^{\alpha}:=\prod_{j=2}^N \left(h(q_1-q_j)-h\ast \widetilde{k}^N_t(q_1)\right)^{\alpha_j}\;. 
\end{equation}
It follows that
\begin{align*}
\mathbb{E}_t\left(\left(H_1(X)-(N-1)h\ast \widetilde{k}^N_t(q_1)\right)^M\right)
=&\mathbb{E}_t\left(\left(\sum_{i=2}^N\big(h(q_i-q_1)-h\ast \widetilde{k}^N_t(q_1)\big)\right)^M\right)
\\
=&\sum_{\alpha\in\mathcal{M}} \mathbb{E}_t(G^{\alpha})\;.
\end{align*} 

Note that $\mathbb{E}_t(G^\alpha)=0$ whenever there exists a $1\leq j\leq N$ such that $\alpha_j=1$. This can be seen be integrating the $j^{th}$ variable first.

Whenever $|\alpha|> M/2$ there has to be at least one index $j$ such that $\alpha_j=1$. 
Thus 
\begin{align}\label{esplit}
\mathbb{E}_t\left(\left(H_1(X)-(N-1)h\ast \widetilde{k}^N_t(q_1)\right)^M\right)=\sum_{|\alpha|\leq M/2} \mathbb{E}_t(G^{\alpha})\;.
\end{align} 

To proceed we shall need the following two formulae: For any two functions $f,g:\mathbb{R}^3\to \mathbb{R}$
\begin{align}
\|f\ast g\|_\infty\leq& \|f\|_{1\wedge \infty}\|g\|_{1\vee\infty}\label{form1}
\end{align}
where 
\begin{align}
\|\cdot\|_{1\wedge \infty}:=\|\cdot\|_1+\|\cdot\|_\infty
\end{align}
and
\begin{align}\label{oneinfty}
\|g\|_{1\vee\infty}:=\inf_{g_1+g_\infty=g}\left\{\|g_1\|_1+\|g_\infty\|_\infty\right\}\;.
\end{align}

Formula (\ref{form1}) can be proven in the following way: For any $g_1+g_\infty=g$ we have using triangle inequality, Young and H\"older
\begin{align*}
\|f\ast g\|_\infty\leq&\|f\ast g_1\|_\infty+\|f\ast g_\infty\|_\infty\leq \|f\|_\infty\|g_1\|_1+\|f\|_1\|g_\infty\|_\infty\\\leq& \|f\|_{1\wedge \infty}(\|g_1\|_1+\|g_\infty\|_\infty)\;.
\end{align*}
Taking the infimum over all possible $g_1$ and $g_\infty$ the formula follows.

Since $\|\widetilde{k}^N_t\|_1 = 1$ and $\|\widetilde{k}^N_t\|_\infty$ is bounded (cf. equation (\ref{cond})) it holds that $$\|h\ast \widetilde{k}^N_t(q_1)\|_\infty\leq C \|h\|_{1\vee\infty}\;.$$
 Now we test the infimum in $\|h\|_{1\vee\infty}$ (see (\ref{oneinfty})) and get  
\begin{align}\label{hnorm}
\|h\|_{1\vee\infty}\leq &\int_{|q|<1}|h(q)|d^3 q+\sup_{|q|\geq1}|h(q)|\nonumber
\\\nonumber\leq&\int_{|q|<1}CN^{-\frac{2\lambda+2}{9}}|q|^{-\lambda}d^ 3q+CN^{\frac{-2\lambda-2}{9}}
\\\leq&CN^{\frac{-2\lambda-2}{9}}\;.
\end{align}
It follows that
\begin{equation}\label{hfalt}\left\|h\ast \widetilde{k}^N_t(q_1)\right\|_\infty\leq C N^{\frac{-2\lambda-2}{9}}\leq CN^{\frac{\lambda-2}{9}}\end{equation}
and thus $\left|h(q_1-q_j)-h\ast \widetilde{k}^N_t(q_1)\right|\leq CN^{\frac{\lambda-2}{9}}$ and
 \begin{align*}
 \left|h(|q_1-q_j|-h\ast \widetilde{k}^N_t(q_1)\right|^n=&\left|h(|q_1-q_j|-h\ast \widetilde{k}^N_t(q_1)\right|^{n-2}\left|h(|q_1-q_j|-h\ast\widetilde{k}^N_t(q_1)\right|^2
 \\\leq& CN^{\frac{(n-2)(\lambda-2)}{9}} \left|h(|q_1-q_j|-h\ast\widetilde{k}^N_t(q_1)\right|^2 \;.
\end{align*}
Using this and  $\mathbb{E}\left[\left(Z-\mathbb{E}[Z]\right)^2 \right]\leq \mathbb{E}[Z^2]$ for any random variable $Z$ one gets  
\begin{align}\label{hterm}
&\left|\int \widetilde{k}^N_t(q_j)\left(h(|q_1-q_j|)-h\ast\widetilde{k}^N_t(q_1)\right)^nd^{3}q_j\right|
\\\nonumber\leq&CN^{\frac{(n-2)(\lambda-2)}{9}} \left|\int \widetilde{k}^N_t(q_j)\left(h(|q_1-q_j|-h\ast \widetilde{k}^N_t(q_1)\right)^2d^{3}q_j\right|
\\\nonumber\leq&CN^{\frac{(n-2)(\lambda-2)}{9}} \left|\int \widetilde{k}^N_t(q_j)\left|h(|q_1-q_j|)\right|^2d^{3}q_j\right|\;.
\end{align}
The expectation value of $\left|h(q_1-q_j)\right|^2$ can be estimated by
\begin{align*}
&\int \widetilde{k}^N_t(q_j)\left|h(q_1-q_j)\right|^2d^3q_j\\\leq&\int_{|q_1-q_j|<N^{-1/3}}  \widetilde{k}^N_t(q_j)\left|h(q_1-q_j)\right|^2d^3q_j+\int_{|q_1-q_j|\geq N^{-1/3}} \widetilde{k}^N_t(q_j)\left|h(q_1-q_j)\right|^2d^3q_j
\\\leq &C\int_{|q|<N^{-1/3}} N^{\frac{2\lambda-4}{9}}d^3q+CN^{-\frac{4\lambda+4}{9}}\int_{|q|\geq N^{-1/3}}  |q|^{-2\lambda} d^3q_j
\\\leq& CN^{\frac{2\lambda-13}{9}}+CN^{-\frac{4\lambda+4}{9}-\frac{1}{3}(3-2\lambda)}
\leq CN^{\frac{2\lambda-13}{9}}\;.
\end{align*}
Since $\lambda>3/2$ (\ref{hterm}) gives
\begin{align*} \left|\int \widetilde{k}^N_t(q_j)\left(h(|q_1-q_j|-h\ast \widetilde{k}^N_t(q_1)\right)^nd^{3}q_j\right|&\leq CN^{\frac{(n-2)(\lambda-2)}{9}}
N^{\frac{2\lambda-13}{9}}
\\&\leq CN^{\frac{n(\lambda-2)}{9}} N^{-1}
\;.
\end{align*}

Using (\ref{defG}) we get that
\begin{align*}
\mathbb{E}_t(G^\alpha)=&\int\left(\prod_{j=2}^N \int \left(h(|q_1-q_j|-h\ast \widetilde{k}^N_t(q_1)\right)^{\alpha_j} k^N_t(x_j)d^6x_j\right) k^N_t(x_1)d^ 6x_1 
\\\leq &\int\left(\prod_{\alpha_j\neq 0} CN^{\frac{\alpha_j(\lambda-2)}{9}} N^{-1}\right) k(x_1)d^6x_1 
\;.
\end{align*}
Recall that the number of $\alpha_j$ which are not equal to zero is $|\alpha|$ and that $\sum_{j=1}^N \alpha_j=M$. It follows that
\begin{align*}
\mathbb{E}_t(G^\alpha)
\leq &\int\left(  C^{|\alpha|} N^{\frac{M(\lambda-2)}{9}} N^{-|\alpha|}\right) k(x_1)d^ 6x_1 
\\=&C^{|\alpha|} N^{\frac{M(\lambda-2)}{9}} N^{-|\alpha|}\;.
\end{align*}

For any $k\in\{1,\ldots,M\}$ the number of multi-indices $\alpha$ with $|\alpha|=k$ can be calculated by simple combinatorics. Any such $\alpha$ can be uniquely identified by giving first the set $\alpha(\{1,2,\ldots,M\})$ and then any surjective map from $\{1,\ldots,M\}$ into this set.

The number of surjective maps is of course smaller than the number of all maps into this set. Thus the number of indices $\alpha$ with $|\alpha|=k$ can be estimated by
$$\sum_{|\alpha|=k}1\leq {N \choose k} M^k\leq N^k M^M\;. $$

It follows with (\ref{esplit}) that 
\begin{align}\nonumber
\mathbb{E}_t\left(\left(H_1(X)-(N-1)h\ast \widetilde{k}^N_t(q_1)\right)^M\right)\leq& \sum_{k\leq M/2} N^k M^M  C^kN^{-k +M\frac{\lambda-2}{9}}
\\\leq& C^M \frac{M}{2} M^{M} N^{M\frac{\lambda-2}{9}}\;.
\end{align} 
Since $\lambda<2$ we can find for any $\gamma>0$ a $M$ and a constant $C_\gamma$ such that the right hand side is smaller than $C_\gamma N^{-\gamma}$. It follows that $\mathbb{P}_t(\D_1)\leq \mathbb{E}_t(H_1^M)\leq C_\gamma N^{-\gamma}$ and we get (\ref{daszeigen}) which proves the Lemma. 
\end{proof}

\begin{cor}\label{largenumbers}
For any $\gamma>0$ there exists a $C_\gamma<\infty$ such that 
\begin{itemize}
\item[(a)] $\mathbb{P}_0(\B_t)\leq C_\gamma N^{-\gamma}$
\item[(b)] $\mathbb{P}_0(\C_t)\leq C_\gamma N^{-\gamma}$
\end{itemize}
\end{cor}

\begin{proof}
First note that $\mathbb{P}_0(\B_t)=\mathbb{P}_t(\Phi^N_{t,0}(\B_t))$ and $\mathbb{P}_0(\C_t)=\mathbb{P}_t(\Phi^N_{t,0}(\C_t))$.  Note, that the funciton $h$ in Lemma \ref{largenumberslem} was defined in such a way that $(N-1)^{-1}N^{\frac{7-2\lambda}{9}}f^{N}$ and $(N-1)^{-1}g$ satisfy the bound assumed for $h$. 

Note that $X\in\Phi^N_{t,0}(\B_t)$ implies that
\begin{align*}
|F(X)-\overline F_t(X)|_\infty&>N^{\frac{2\lambda-7}{9}}\\
\text{respectively }\;\;\;N^{\frac{7-2\lambda}{9}}|F(X)-\overline F_t(X)|_\infty&>1\;.
\end{align*}
Correspondingly, $X\in\Phi^N_{t,0}(\C_t)$ implies that 
\begin{align*}
|G(X)-\overline G_t(X)|_\infty&>1\;.
\end{align*}
This is satisfied only if there exists a $j$ such that
$$\left|G_j(X)-g\ast \widetilde k^N_t(g_j)\right|=\left|\sum_{i\neq j}(N-1)^{-1}g(q_j-q_i)-g\ast \widetilde k^N_t(q_j)\right|>1\;.$$
Choosing $h=(N-1)^{-1}g$ this reads
$$\left|\sum_{i\neq j}h(q_j-q_i)-(N-1)h\ast \widetilde k^N_t(x)\right|>1\;.$$
 It follows that
 the sets $\Phi^N_{t,0}(\B_t)$ and $\Phi^N_{t,0}(\C_t)$ are subsets of the set $\D$ defined in Lemma \ref{largenumberslem}. Thus one can find for any $\gamma>0$ a $C_\gamma<\infty$ such that
\begin{align*}
\mathbb{P}_0(\B_t)=\mathbb{P}_t(\Phi^N_{t,0}(\B_t))\leq C_\gamma N^{-\gamma}\;,
\\\mathbb{P}_0(\C_t)=\mathbb{P}_t(\Phi^N_{t,0}(\C_t))\leq C_\gamma N^{-\gamma}\;.
\end{align*}

\end{proof}

\begin{lem}\label{lipschitz}
$\left|V(\Psi^N_t(X))-\overline V_t(\Phi^N_t(X))\right|_\infty\leq CJ_t(X)N^{-1/3}+ N^{\frac{2\lambda-7}{9}}$ for all $X\in (\A_t\cup \B_t\cup \C_t)^c$.
\end{lem}

\begin{proof}
Let $X\in (\A_t\cup \B_t\cup \C_t)^c$, $Y:=\Psi^N_{t,0}(X)$ and $Z:=\Phi^N_{t,0}(X)$.

The difference $\left|V(Y)-\overline V_t(Z)\right|_\infty$ comes from a difference in the respective forces and a difference in the respective momenta. The latter is bounded by $|Y-Z|_\infty$ and we get 
$$\left|V(Y)-\overline V_t(Z)\right|_\infty\leq\left|F(Y)-\overline F_t(Z)\right|_\infty+|Y-Z|_\infty$$

By triangle inequality
\begin{align}
\left|F(Y)-\overline F_t(Z)\right|_\infty 
\leq&\left|F(Y)- F(Z)\right|_\infty\label{tri1}\\& +\left|F(Z)-\overline F_t(Z)\right|_\infty \label{tri2}\;.
\end{align}
Since $X\notin \B_t$ it follows that 
$\left|F(Z)-\overline F_t(Z)\right|_\infty\leq N^{\frac{2\lambda-7}{9}}$ which controls (\ref{tri2}).

With triangle inequality we get that for any $1\leq j \leq N$ 
\begin{align}
|(F(Y)- F(Z))_j|_\infty=&\left| (N -1)^{-1} 
\sum_{k\neq j}f^{N}(y_j-y_k)-f^{N}(z_j-z_k) 
\right|_\infty
\\&\leq (N -1)^{-1}\sum_{k\neq j}\left|
f^{N}(y_j-y_k)-f^{N}(z_j-z_k)  
\right|_\infty\;.
\end{align}
Since $X\notin \A_t$ it follows that $\left|Y-Z\right|_\infty< N^{-\frac{1}{3}}$.
In particular $|y_j-z_j|_\infty< N^{-\frac{1}{3}}$ and $|y_k-z_k|_\infty< N^{-\frac{1}{3}}$.

Thus Lemma \ref{gkraft} implies
$$|f^{N}(y_j-y_k)-f^{N}(z_j-z_k)|_\infty\leq g (z_j-z_k)|y_j-y_k-z_j+z_k|_\infty\;,$$
and consequently

\begin{align}
|(F(Y)- F(Z))_j|_\infty\leq& (N -1)^{-1}\sum_{k\neq j}g (z_j-z_k)|y_j-y_k-z_j+z_k|_\infty  
\\
\leq&
\left(G(Z)\right)_j 2|Y-Z|_\infty\;.
\end{align}

Since $X\notin \C_t$ it follows that $\left|\left(G(Z)\right)_j\right|\leq \|g\ast k^N_t\|_\infty+1\leq C$ and thus
\begin{align}
|(F(Y)- F(Z))_j|_\infty
\leq&
C |Y-Z|_\infty\;.
\end{align}
Since this holds true for any $1\leq j \leq N$ and furthermore $X\notin \A_t$, we obtain

\begin{align}
|F(Y)- F(Z)|_\infty + |Y-Z|_\infty
\leq&
C J_t(X)N^{-1/3}
\end{align}
and the Lemma follows.
\end{proof}

\paragraph{Proof of Lemma \ref{mainlem}}

With Corollary \ref{largenumbers} and Lemma \ref{lipschitz} we can control (\ref{line1}) and (\ref{line2}). It follows that there exists a $C>0$ such that
\begin{align}
\mathbb{E}_0(J_{t+dt}-J_{t}) \leq& 2CN^{-1}dt
+\mathbb{E}_0(C (J_t(X)N^{-1/3}+ N^{\frac{2\lambda-7}{9}})) N^{1/3}dt\nonumber+o_N(dt)
\\\leq&2CN^{-1}dt+C\mathbb{E}_0(J_t(X))dt+N^{\frac{2\lambda-4}{9}}dt+o_N(dt)\;.
\end{align}

Using Gronwall's Lemma and that $\frac{2\lambda-4}{9}$ is negative we get Lemma \ref{mainlem}.

\section{Proof of Theorem \ref{theo2}}

Let for any $N\in\mathbb{N}\cup\{\infty\}$ \begin{align*}
x_t^N(q_0,p_0):=&(q_t^N(q_0,p_0),p_t^N(q_0,p_0)):= \varphi^N_{0,t}(q_0,p_0)\\
\beta_t:=&\sup_{q_0,p_0\in\mathbb{R}^3}|x_t^N(q_0,p_0)-x_t^\infty(q_0,p_0)|
\end{align*}

Note that by (\ref{cond2}) the Vlasov-force is bounded for all times. It follows that there exists a  time-dependent $C<\infty$ uniform in $p_0$ and $q_0$ such that 
\begin{align}
\label{propbound1}p_t^\infty-p_0< &C \\
\label{propbound2}q_t^\infty-p_0 t< &C\;.
\end{align}

We shall now estimate $\beta_t$ via Gronwall's Lemma:
\begin{align*}
\partial_t \beta_t\leq&\sup_{q_0,p_0\in\mathbb{R}^3}|\left(p_t^N(q_0,p_0)-p_t^\infty(q_0,p_0),\widetilde{k}^N_t\ast f^N(q_t^N(q_0,p_0))-\widetilde{k}^\infty_t\ast f^\infty(q_t^\infty(q_0,p_0))\right)| 
\\\leq& \beta_t+
\sup_{q_0,p_0\in\mathbb{R}^3}|\widetilde{k}^N_t\ast f^N(q_t^N(q_0,p_0))-\widetilde{k}^N_t\ast f^N(q_t^\infty(q_0,p_0))| 
\\&+\sup_{q_0,p_0\in\mathbb{R}^3}|\widetilde{k}^N_t\ast f^N(q_t^\infty(q_0,p_0))-\widetilde{k}^N_t\ast f^\infty(q_t^\infty(q_0,p_0))| 
\\&+\sup_{q_0,p_0\in\mathbb{R}^3}|\widetilde{k}^N_t\ast f^\infty(q_t^\infty(q_0,p_0))-\widetilde{k}^\infty_t\ast f^\infty(q_t^\infty(q_0,p_0))|\;. 
\\
\leq& \beta_t+\|\widetilde{k}^N_t\ast f^N\|_L\|q_t^N-q_t^\infty\|_\infty
\\&+\|\widetilde{k}^N_t\ast f^N-\widetilde{k}^N_t\ast f^\infty\|_\infty
\\&+\|\widetilde{k}^N_t\ast f^\infty-\widetilde{k}^\infty_t\ast f^\infty\|_\infty
\\\leq&
\beta_t+\|\widetilde{k}^N_t\|_\infty\|\nabla f^N\|_1\beta_t
+\|\widetilde{k}^N_t\|_\infty\|f^N-f^\infty\|_1 
\\&+\|\widetilde{k}^N_t-\widetilde{k}^\infty_t\|_\infty\|f^\infty\|_1\;. 
\end{align*}

Let us assume that $t$ is such that $\beta_t\leq 1$. Then 
\begin{align*}\left|\widetilde{k}^N_t(q)-\widetilde{k}^\infty_t(q)\right|
\leq& \int \left|k_0(q^N_t,p^N_t)-k_0(q^\infty_t,p^\infty_t)\right| d^3p_0
\\\leq&\int \left(\sup_{h\in\mathbb{R}^3;|h|=1}\nabla k_0(q^\infty_t,p^\infty_t+h)\right)|x^N_t-x^\infty_t| d^3p_0 
\;.
\end{align*}

By (\ref{propbound1}), (\ref{propbound2}), and the assumptions of the Theorem (i.e.  $k_0(x)\leq C_0(1+|x|)^{-7}$) we get that  $\int \left(\sup_{h\in\mathbb{R}^3;|h|=1}\nabla k_0(q^\infty_t,p^\infty_t+h)\right) d^3p_0 $   is bounded,
and hence  \begin{align*}\|\widetilde{k}^N_t-\widetilde{k}^\infty_t\|_\infty 
\leq& C \beta_t\;.
\end{align*}

Thus 
for times $t$ with $\beta_t\leq 1$
\begin{align*}
\partial_t \beta_t
\leq&  C\beta_t
+C\|f^N-f^\infty\|_1 \;.
\end{align*}

Since $ \|f^N-f^\infty\|_1\leq N^{-1/3}$ we get with Gronwall's Lemma that for any $t\geq0$ there exists a $C$ such that for any $0\leq s\leq t$
$$\sup_{q_0,p_0\in\mathbb{R}^3}|x_s^N(q_0,p_0)-x_s^\infty(q_0,p_0)|\leq C N^{-1/3}\;,$$
which implies \begin{equation}
\label{formel}
\left\|\Phi^N_{s,0}-\Phi^\infty_{s,0}\right\|_\infty<CN^{-1/3}\;.
\end{equation}
Together with Theorem \ref{theo1}, this completes the proof.
\eop

\section{Proof of Corollary \ref{cor1}}

We present the proof of the Corollary only under the conditions of Theorem \ref{theo1}. The proof under the conditions of Theorem \ref{theo2} is equivalent. 

Let $\M_t\subset \mathbb{R}^{6N}$ be given by
$$X\in\M\Leftrightarrow \left|\Psi^N_{t,0}(X)-\Phi^N_{t,0}(X)\right|_\infty>N^{-1/3}$$ 

From Theorem \ref{theo1} we get that 
\begin{equation}
\label{conv}\mathbb{P}_0(\M)\leq C_1N^{\frac{2\lambda-4}{9}}\;.
\end{equation}

Note that

\begin{align}
&d_L(\F^{(s)}_t,\left(k^N_t)^{\otimes s}\right)=\sup_{g\in\mathcal{L}}\left|\int\left(\F_t(X)-\prod_{j=1}^N k^N_t(x_j)\right) g(x_1, \dots, x_s)d^{6N}X\right| \nonumber
\\&=\sup_{g\in\mathcal{L}}\left|\int\left(\F_0(\Psi^N_{0,t}(X))-\prod_{j=1}^N k_0(\varphi^N_{0,t}(x_j))\right) g(x_1, \dots, x_s)d^{6N}X\right|\nonumber
\\&=\sup_{g\in\mathcal{L}}\left|\int\left(\F_0(\Psi^N_{0,t}(X))-\F_0(\Phi^N_{0,t}(X))\right) g(x_1, \dots, x_s)d^{6N}X\right|\nonumber
\\&=\sup_{g\in\mathcal{L}}\left|\int \F_0(X) \left(g\left(\left(\Psi^N_{t,0}(X)\right)_1, \dots , \left(\Psi^N_{t,0}(X)\right)_s\right)-g\left(\left(\Phi^N_{t,0}(X)\right)_1, \dots , \left(\Phi^N_{t,0}(X)\right)_s\right)\right)d^{6N}X\right|\nonumber
\\&=\sup_{g\in\mathcal{L}}\left|\int_{\M_t} \F_0(X) \left(g\left(\left(\Psi^N_{t,0}(X)\right)_1, \dots , \left(\Psi^N_{t,0}(X)\right)_s\right)-g\left(\left(\Phi^N_{t,0}(X)\right)_1, \dots , \left(\Phi^N_{t,0}(X)\right)_s\right)\right)d^{6N}X\right|\label{erste}
\\&+\sup_{g\in\mathcal{L}}\left|\int_{\M_t^c} \F_0(X) \left(g\left(\left(\Psi^N_{t,0}(X)\right)_1, \dots , \left(\Psi^N_{t,0}(X)\right)_s\right)-g\left(\left(\Phi^N_{t,0}(X)\right)_1, \dots , \left(\Phi^N_{t,0}(X)\right)_s\right)\right)d^{6N}X\right|\label{zweite}\;.
\end{align}
From (\ref{conv}) and since $\|g\|_\infty=1$ it follows that (\ref{erste}) tends to zero as $N\to\infty$.

Using that $\|g\|_L=1$ we obtain
$$ \sup_{X\in\M^c}\left\{ g\left(\left(\Psi^N_{t,0}(X)\right)_1, \dots , \left(\Psi^N_{t,0}(X)\right)_s\right)-g\left(\left(\Phi^N_{t,0}(X)\right)_1, \dots , \left(\Phi^N_{t,0}(X)\right)_s\right)\right\}\leq N^{-1/3}\;.$$

Hence $(\ref{zweite})\leq N^{-1/3}$ and $d_L(\F^{(s)}_t,\left(k^N_t)^{\otimes s}\right)$ converges to zero as $N\to\infty$.  Since $\lambda > 3/2$ the rate of convergence is  given by $d_L(\F^{(s)}_t,\left(k^N_t)^{\otimes s}\right)\leq C N^{\frac{2\lambda-4}{9}} $ for some time-dependent, finite $C$.

\eop

\paragraph{Acknowledgments:}

We originally intended to publish this paper in a special issue dedicated to the $60^{\text{th}}$ anniversary of Herbert Spohn. Unfortunately, substantial corrections delayed the completion of the mansucript. We wish to thank Herbert for his friendship and his support. Furthermore, very helpful discussions with Detlef D\"urr and Dustin Lazarovici are gratefully acknowledged. Finally, we would like to thank the referees for carefully proofreading our paper. Their detailed comments helped us to substantially improve our manuscript.

\bibliography{meanfield}
\bibliographystyle{apalike}

\end{document}